%% file: main.tex
\documentclass{article}

\usepackage[english]{babel}

\usepackage[letterpaper,top=2cm,bottom=2cm,left=3cm,right=3cm,marginparwidth=1.75cm]{geometry}

\usepackage{amsmath}
\usepackage{graphicx}
\usepackage{multirow}
\usepackage{subfigure}
\usepackage{xcolor}
\usepackage[hyphens]{url}
\usepackage{hyperref}
\usepackage{caption}
\usepackage{cite}
\usepackage{authblk}
\usepackage{wrapfig} 
\usepackage{booktabs}

\title{Development of A Real-time POCUS Image Quality Assessment and Acquisition Guidance System}

\author[1]{Zhenge Jia}
\author[1]{Yiyu Shi}
\author[2]{Jingtong Hu}
\author[3]{Lei Yang}
\author[4]{Benjamin Nti}

\affil[1]{Department of Computer Science and Engineering, University of Notre Dame, USA}
\affil[2]{Department of Electrical and Computer Engineering, University of Pittsburgh, USA}
\affil[3]{Department of Information Sciences and Technology, George Mason University, USA}
\affil[4]{Clinical Emergency Medicine and Pediatrics, Indiana
University School of Medicine, USA}

\date{}
\begin{document}
\maketitle

\begin{abstract}
Point-of-care ultrasound (POCUS) is one of the most commonly applied tools for cardiac function imaging in the clinical routine of the emergency department and pediatric intensive care unit. 
The prior studies demonstrate that AI-assisted software can guide nurses or novices without prior sonography experience to acquire POCUS by recognizing the interest region, assessing image quality, and providing instructions~\cite{nti2022artificial, cap-cheema2021artificial, cap-narang2021utility, cap-schneider2021machine, butter-baribeau2020handheld}. 
However, these AI algorithms cannot simply replace the role of skilled sonographers in acquiring diagnostic-quality POCUS. 
Unlike chest X-ray, CT, and MRI, which have standardized imaging protocols, POCUS can be acquired with high inter-observer variability. 
Though being with variability, they are usually all clinically acceptable and interpretable. 
In challenging clinical environments, sonographers employ novel heuristics to acquire POCUS in complex scenarios. 
To help novice learners to expedite the training process while reducing the dependency on experienced sonographers in the curriculum implementation, 
We will develop a framework to perform real-time AI-assisted quality assessment and probe position guidance to provide training process for novice learners with less manual intervention.
\end{abstract}

\input{latex/intro.tex}
\input{latex/innovation.tex}
\input{latex/results.tex}

\bibliographystyle{IEEEtran}
\bibliography{sample}

\end{document}

%% file: latex/intro.tex
\section{Introduction}

Automated acquisition guidance can help novice learners to study how to manipulate probes to acquire high-quality POCUS images.
However, the difference between human vision and machine vision may confuse novice learners, because the guidance from sonographers and the DNN model may be inconsistent due to the different interpretations of images between the DNN-based models and humans~\cite{zeng2021segmentation}.
For example, DNNs rely more on textures and high-frequency features rather than shapes and low-frequency features. Our previous works show that machine learning based detection methods on medical image can achieve significantly high accuracy~\cite{xu2019whole,wang2019msu,wang2020ica,xu2020imagechd,wang2021echocp,lu2022rt,zeng2008cochlear,wu2022fairprune,zeng2021hardware}.
This means images considered high-quality for humans may not be high-quality for DNN-based machine vision.
The discrepancy becomes more significant due to the large inter-observer variability. 
As a result, if the acquired images are of high quality for both human and DNN-based segmentation models, the segmentation accuracy can be improved, and accordingly, the human efforts needed to correct or identify the bad segmentation can be reduced. 
As an example in a closely-related area, previous works for medical image compression demonstrated that image compressed with machine vision considered achieved significantly higher segmentation accuracy than conventional methods that only consider human vision at the same compression rate~\cite{liu2019machine}. 
Therefore, we will explore methods to optimize the image quality assessment and acquisition guidance to be preferred by both human and machine models.

%% file: latex/innovation.tex
\section{Methods}

\begin{figure}[!htb]
  \centering
  \includegraphics[width=1\textwidth]{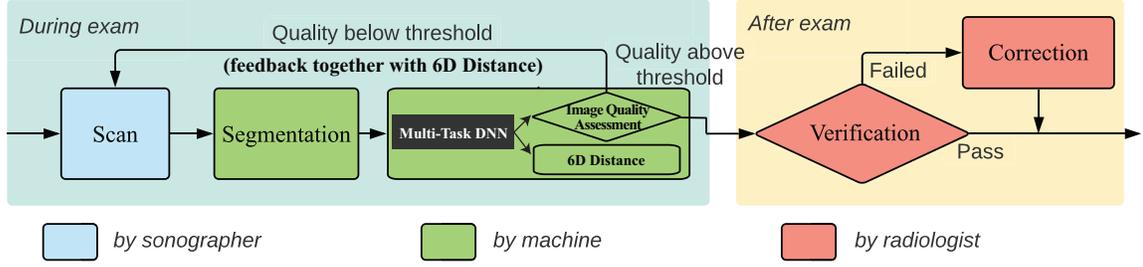}
  \caption{Framework of simultaneously optimizing image quality assessment and acquisition guidance by a multi-task DNN.}
  \label{fig:data_acq}
\end{figure}

The overall framework proposed is shown in Figure~\ref{fig:data_acq}. 
During data acquisition, the scan is checked by sonographers to ensure high visual quality for humans.
After that, the segmentation model will produce segmentation results. 
Then a multi-task DNN will predict the quality of the input scan image according to the segmentation results, and meanwhile, the model will generate 6D (probe position plus orientation) geometric distances, which can be feedback to novice learners for guidance on how to move from the current probe location and the probe location anticipated to optimize the image. 
Here, such a DNN model needs to be conducted in a real-time manner, because the learners need immediate feedback from the model to learn how to move the probe.
If the predicted quality is lower than the predefined threshold, the novice learner will be prompted to do a re-scan with guidance. 
The study/exam process is done once the predicted image quality exceeds the threshold (a max number of re-scans can be set to avoid too many trials). 
Because the prediction model helps learners to select the data that the model can segment well, the quality of the segmentation results and the subsequent medical analysis will be improved. 

\begin{wraptable}[12]{R}{0.4\textwidth} 
\begin{center}
\vspace{-22pt}
\resizebox{0.4\textwidth}{!}{
\begin{tabular}{p{8.cm}r}
\toprule
\textbf{Box 1. Equations} \\
$C=c_c\int \alpha f(\alpha)d\alpha$; & Eq.1 \\
$\diamond$ $C$: Original average cost & \\
$\diamond$ $c_c$: Average cost of correcting a failed segmentation & \\
$\diamond$ $\alpha$: Probability of segmentation fails & \\
$\diamond$ $f(\alpha)$: Probability density function of $\alpha$ & \\
$C'_{\alpha} = \alpha(1-r)c_c+\frac{\alpha r}{p}(c_s+C'_{\alpha})$; & Eq.2 \\
$\diamond$ $C'$: New average cost & \\
$\diamond$ $c_s$: Average cost of an additional scan & \\
$\diamond$ $p$: Precision of the quality prediction model & \\
$\diamond$ $r$: Recall of the quality prediction model & \\
$C'_{\alpha} = \frac{p\alpha c_c-p\alpha rc_c+\alpha r c_s}{p-\alpha r}$; & Eq.3 \\
$h(\alpha)=\frac{C'_{\alpha}}{C_{\alpha}} = \frac{p-pr+r\frac{c_s}{c_c}}{p-\alpha r}$; & Eq.4 \\
$\diamond$ $h(\alpha)$: Ratio of new cost to original cost for $\alpha$ & \\
$\frac{C'}{C}=\frac{\int \alpha f(\alpha) h(\alpha) d\alpha}{\int \alpha f(\alpha) d\alpha}$; & Eq.5 \\
\bottomrule
\end{tabular}}
	\end{center}
	\label{tab:equations_aim2}
\end{wraptable}

The key component of the framework is the multi-task DNN model for both image quality prediction and 6D distance generation. 
We will develop a DNN with a shared encoder (i.e., the front layers of DNN) to extract common features, and then it will be split into two paths for image quality assessment and acquisition guidance tasks, respectively.
For the image quality assessment task, we will use a modification of DNN-based segmentation model as the image quality prediction model. It can take advantage of existing techniques such as model architecture for image segmentation, uncertainty estimation of neural networks, and the attention mechanism. We will compare the quality prediction models that are trained to predict conventional quality metrics and the new quality metric proposed in Aim 1. In this way, we can additionally validate whether the new metric also helps in this framework.
On the other hand, for the 6D distance generation task, we will add a tail after the encoder, and each output neuron will correspond to one distance in the 6D distance.
We will track the image quality after the learners move the probe according to the predicted 6D distance to evaluate the effectiveness of the model.

A threshold applied to the predicted image quality will control the learning loop.
A low threshold means a re-scan is suggested only if the predicted image quality is quite low and thus the feedback information to novice learners is minimal. A high threshold means the data is satisfying only if the predicted quality is quite high which leads to higher image quality, and in turn, it indicates a strict requirement for learners. 
In practice, the threshold can be set based on the progress of the learner. 

The learning quality of a learner followed the proposed framework can be analyzed via a theoretical cost model.
Because the cost for the initial scan and verification are not affected by the proposed method, we focus on the cost of correcting segmentation and the cost of doing a re-scan in the analysis below. We consider the segmentation on the video obtained fails if correction is needed, and denote $\alpha$ as the failure probability. 
A fixed $\alpha$ for all subjects may be inaccurate because each subject may have a unique condition that makes it easier or more difficult to obtain images that can be segmented well. Therefore, we consider each subject has a unique $\alpha$ and repetitive scans for the same subject are independent. 
The original average per-subject cost is given in Eq. 1 (Box 1).
The quality prediction model is used to identify failed scans. The new average cost for a certain $\alpha$ is given in Eq. 2. 
The first term is for predicted passed segmentation and the second term is for predicted failed segmentation. Solving Eq. 2 we get Eq. 3.
It can be derived that, for a certain $\alpha$, the new average cost is lower than the original average cost if and only if $p>\alpha+\frac{c_s}{c_c}$. This indicates a lower bound of precision can reduce the cost. 
Finally the ratio of the new cost $C'$ to the original cost $C$ is given in Eq. 4 and Eq. 5.

%% file: latex/results.tex
\section{Results}

\begin{table}[!htb] 
\vspace{-0pt}
	\caption{Estimated cost reduction. $\alpha$ is the probability of a segmentation fails. $\frac{c_s}{c_c}$ is the ratio of the scan cost to the correction cost. Cost reduction is $1-\frac{C'}{C}$.}
	\begin{center}
\resizebox{0.7\textwidth}{!}{
\begin{tabular}{lcccccccccccc}
 \toprule
 $\alpha$       & & 0.2 & 0.3 & 0.2 & 0.2 & 0.2 & 0.2 \\
  $\frac{c_s}{c_c}$& & 0.1 & 0.1 & 0.2 & 0.1 & 0.1 & 0.1 \\
 Precision & & 0.8 & 0.8 & 0.8 & 0.6 & 0.9 & 0.7 \\
Recall     & & 0.8 & 0.8 & 0.8 & 0.6 & 0.7 & 0.9 \\
\midrule
Cost reduction   & & 64\% & 57\%  &  50\% & 37\% & 55\% & 69\% \\
 \bottomrule
 \end{tabular}}
	\end{center}
	\label{tab:example}
\end{table}

This cost model reflects how various factors affect the cost-saving of the proposed framework. $f(\alpha)$ accounts for the data distribution in the specific applications which can be estimated by examining the segmentation results of interest. Bigger $\alpha$ leads to lower cost reduction because it takes more scans to obtain good segmentation. $h(\alpha)$ is determined by $\frac{c_s}{c_c}$, $p$, and $r$. $\frac{c_s}{c_c}$ is the ratio of the re-scan cost to the correction cost which can be easily estimated by sonographers and radiologists. Smaller $\frac{c_s}{c_c}$ means smaller $\frac{C'}{C}$ and thus more cost reduction. An accurate prediction model is crucial for the cost reduction as reflected by $p$ and $r$. In Table~\ref{tab:example}, we provide a few numerical examples of the cost reduction when variables are set to typical values. The cost model shows a substantial possible cost reduction.

\section{Conclusions}
While we can measure the learning quality by quantitative metrics, it is also reasonable to assume that the final segmentation results are equivalently satisfactory because they are all verified by radiologists~\cite{zeng2021segmentation}. Therefore, an alternative goal is to reduce the manual correction effort needed. 
In this paper, we approach this goal by using the framework in Figure~\ref{fig:data_acq} to predict the human effort needed to correct the automatic segmentation results rather than the image quality metrics. In this way, we aim to directly reduce the human effort needed in the whole segmentation process.